\documentclass[prl,reprint,amsmath,amssymb,aps,superscriptaddress]{revtex4-1}
\pdfoutput=1
\usepackage{geometry,enumerate,amsmath,amssymb}
\usepackage{graphicx}
\usepackage{fullpage}

\newcommand{\bnabla}{\mbox{\boldmath $\nabla$}}

\begin{document}
\title{Colonies of threaded rings in excitable media}
\author{Fabian Maucher}
\affiliation{Department of Physics and Astronomy, Aarhus University, Ny Munkegade 120, DK 8000 Aarhus, Denmark}
\affiliation{Department of Mathematical Sciences, Durham University, Durham DH1 3LE, United Kingdom}
\affiliation{Joint Quantum Centre (JQC) Durham-Newcastle, Department of Physics, Durham University, Durham DH1 3LE, United Kingdom}
\author{Paul Sutcliffe}
\affiliation{Department of Mathematical Sciences, Durham University, Durham DH1 3LE, United Kingdom}
\date{June 2020}
\begin{abstract}

A thring is a recent addition to the zoo of spiral wave phenomena found in excitable media and consists of a scroll ring that is threaded by a pair of counter-rotating scroll waves. This arrangement behaves like a particle that swims through the medium.
Here, we present the first results on the dynamics, interaction and collective behaviour of several thrings via numerical simulation of the reaction-diffusion equations that model thrings created in chemical experiments.
We reveal an attraction between two thrings that leads to a stable bound pair that thwarts their individual locomotion. Furthermore, such a pair emits waves at a higher frequency than a single thring, which protects the pair from the advances of any other thring and rules out the formation of a triplet bound state.
As a result, the long-term evolution of a colony of thrings ultimately yields an unusual frozen nonequilibrium state consisting of a collection of pairs accompanied by isolated thrings that are inhibited from further motion by the waves emanating from the pairs.
\end{abstract}

\maketitle

The creation of stable bound pairs of particle-like structures is fundamental to a range of natural systems and can give rise to qualitatively new behaviour. 
Prominent examples include composite bosons, such as the formation of Cooper pairs that are foundational for superconductivity~\cite{Cooper:PhysRev:1956}, and excitons, which are bound states of an electron and an electron hole, that transport energy without a flux of net charge~\cite{Frenkel:PhysRev:1931}.   
These, and many other examples, share the common feature that there is an energy landscape underlying pair creation. In nonequilibrium systems the formation of stable pairs appears to be much less likely, as there is no energy landscape minimum to drive the formation of a bound state. 
Here, we present an example of nonequilibrium dynamics for particle-like objects composed of spiral waves in excitable media and show that these can form bound structures. Furthermore, these bound states display different properties to their individual counterparts, producing an unusual frozen state for the collective behaviour of a colony.

Spiral waves in excitable media occur in a range of biological and chemical settings, including the Belousov-Zhabotinsky (BZ) reaction~\cite{Epstein:book:1998}, ventricular fibrillation~\cite{Witkowski:Nature:1998}, and 
chemotaxis in Dictyostelium~\cite{Kessler:PRE:1993,Sawai:Nature:2005}.
In a three-dimensional medium, spiral waves become extended scroll waves with a line-like filament core. The filament must either end on the boundary of the medium or form a closed loop, with the simplest example being a circular scroll ring. Filament dynamics has a complicated dependence on local properties, such as curvature and twist~\cite{Keener:PhysicaD:1988,Biktashev:1994}, but even more elusive is the influence of global topological properties on filament motion. Numerical simulations reveal elaborate evolutions for filaments that are knotted or linked~\cite{Sutcliffe:PRE:2003,Sutcliffe:PRL:2016,Sutcliffe:PRE:2017,Sutcliffe:JPhysA:2018,Sutcliffe:Nonlinearity:2019,Gareth:PRE:2019},
including knot untangling without untying. These results make it apparent that an important mechanism that drives filament dynamics is the location of wavefront collision interfaces, but these evolve in a subtle manner due to small differences in the frequencies of wave emissions from differing parts of a filament. As a result, there is currently no effective analytic prescription to predict filament motion in topologically complex settings, even at a qualitative level.

Chemical experiments realizing isolated scroll rings in excitable media have been conducted with remarkable precision~\cite{Welsh:Nature:1983,Bansagi:PRL:2006,Azhand:EPL:2014,Steinbock:NJP:2015}. 
However, experimental access to knotted filaments remains an elusive goal. Recently, the first controlled experiments on linked filaments were performed,
by creating a scroll ring that is threaded by a pair of counter-rotating scroll waves that end on the boundary of a thin excitable medium~\cite{Cincotti:arxiv:2019}. Such an arrangement has been named a thring, and was achieved in a photosensitive version of the BZ reaction through the application of a certain optical spatio-temporal protocol. Numerical studies~\cite{Cincotti:arxiv:2019} have shown that thrings in thin excitable media are self-propelling particle-like objects, where the topology induces a swimming-type motion in the plane of the scroll ring that is parallel to a boundary of the excitable medium.

There is no fundamental problem in extending the previously mentioned light templating protocol, used to create an isolated thring~\cite{Cincotti:arxiv:2019}, to the case of many thrings.
Hence, studying their collective behaviour should be as amenable to experiment as the case of a single thring. Motivated by the possibility of future experimental investigations and the prospect of gaining further insight into wave interactions with complex topology in excitable media more generally, we perform numerical simulations to study the dynamics, interaction and collective behaviour of several thrings in this relatively simple system. 

To simulate the dynamics of the photosensitive version of the Belousov-Zhabotinsky medium we use the modified Oregonator reaction-diffusion equations~\cite{Krug:JPhysChem:1990} that have already been applied to describe experiments on scroll rings~\cite{Azhand:EPL:2014} and thrings~\cite{Cincotti:arxiv:2019}. 
In dimensionless form these equations read
\begin{align}
\frac{\partial u}{\partial t}&=\frac{1}{\varepsilon}\big[(u(1-u)+w(\beta-u)\big]+\nabla^2 u, \
\frac{\partial v}{\partial t}=u-v, \nonumber \\
\frac{\partial w}{\partial t}&=\frac{1}{\varepsilon'}\big[\Phi + \gamma v-w(\beta+u)\big]+\delta \nabla^2 w,
\label{pde}
\end{align}
where the variables $u,v,w$ are proportional to the concentrations of bromous acid, the oxidized form of the ruthenium catalyst, and bromide ions, respectively. We use the same parameter values as in~\cite{Azhand:EPL:2014,Cincotti:arxiv:2019} that reproduce the chemical experiments, namely, $\varepsilon=0.125, \ \varepsilon'=0.00139, \ \beta=0.002, \ \gamma=1.16, \ \delta=1.12$ and  $\Phi=0.013.$ The parameter $\Phi$ is proportional to the light intensity and the above value represents the light supplied by a projector typically used in experiments.

With the above parameter values the system supports a spiral scroll wave with a wavelength $\lambda=18.6$ and a period $T=6.5.$ To compare these dimensionless values to chemical experiments, $\lambda/2$ and $T/2$ are around a millimetre and a minute respectively. To match to previous experiments on isolated thrings, we consider a thin medium with a thickness $\lambda/2.$
The numerical scheme to solve equations (\ref{pde}) has been described in detail previously~\cite{Cincotti:arxiv:2019}, together with the method to display the results. Briefly, an explicit fourth-order Runge-Kutta method is employed with a timestep $dt=0.005$ and a finite difference approximation for the Laplacian using a 27 point stencil and a lattice spacing $dx=0.5$ on a rectangular grid with no-flux (Neumann) boundary conditions imposed at all boundaries of the medium.
The oxidation of the catalyst is visualized by displaying a heat map of the average value of $v$ along the thin direction of the medium (taken to be the $z$-direction). The filaments are indicated by identifying the regions where the vorticity is localized, determined by the condition $|\bnabla u \times \bnabla v|>0.008.$
\begin{figure}[ht]
\includegraphics[width=\columnwidth]{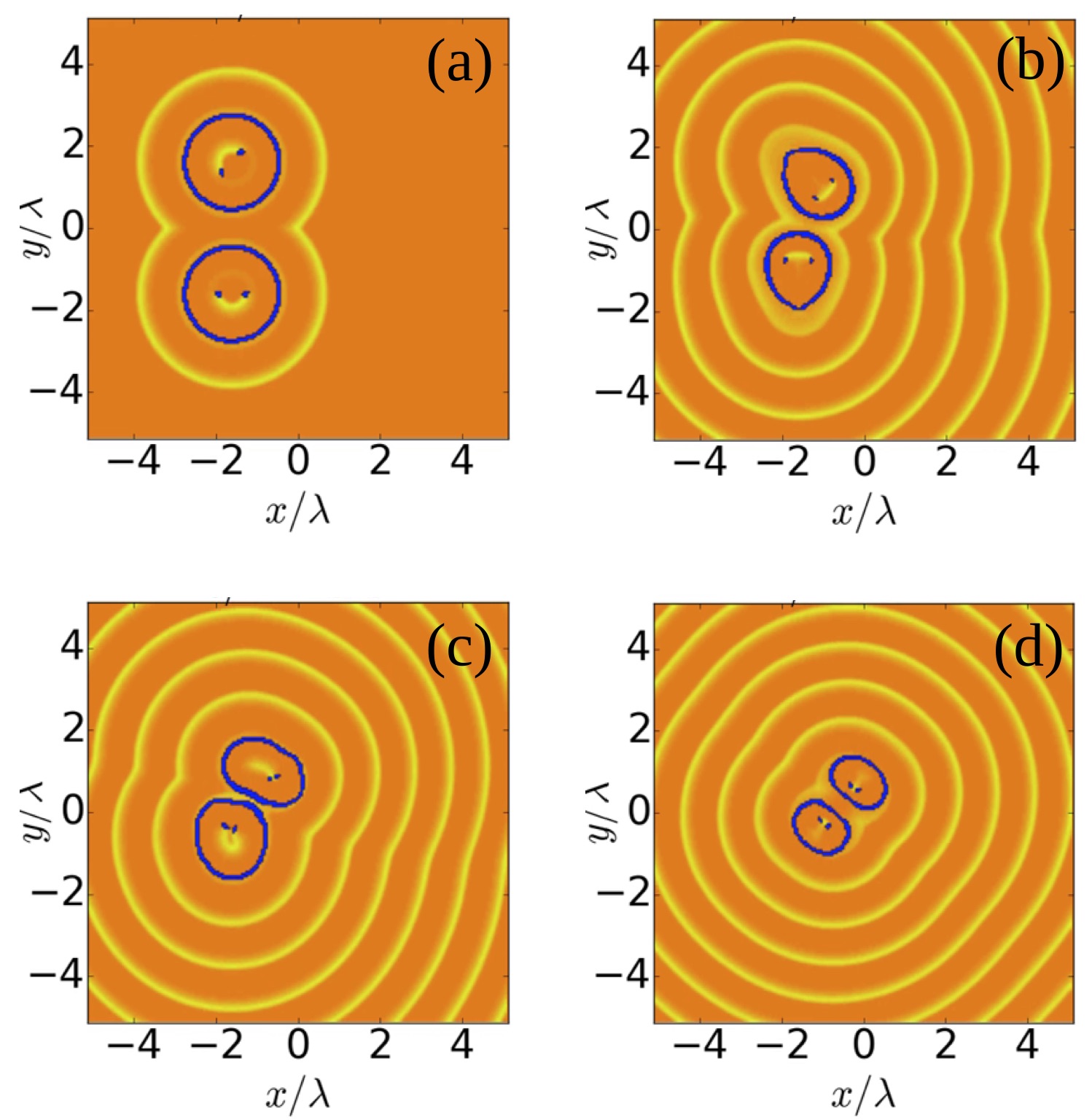}
\caption{The interaction of a pair of thrings, with chemical waves displayed using a heat map of $v$ and filaments shown in blue. (a) The situation shortly after  simulated optical initialization ($t/T=2.5$), (b-c) the collision ($t/T=15.8, \ 23.5$), (d) the long-term dynamics, which involves reorientation and the subsequent formation of a stable bound pair ($t/T=92.4$). 
\label{fig:bound_pair}
}
\end{figure}

To study the interaction of a pair of thrings, we numerically imitate the experimental light templating protocol introduced recently~\cite{Cincotti:arxiv:2019}, to create the pair of thrings depicted in Fig.~\ref{fig:bound_pair}(a). The lower thring is initialized to move along a line parallel to the $y$-axis (in the direction of increasing $y$) and the upper thring initially moves parallel to the diagonal line $y=-x$ (in the direction of decreasing $y$) to yield a glancing collision. In the early stages of the motion, Fig.~\ref{fig:bound_pair}(b), the pair of thrings swim independently towards the anticipated glancing collision. During the collision, there is a deformation of the shape of each thring as they remain glued together, Fig.~\ref{fig:bound_pair}(c), under the influence of an attractive force.
During a long time-span the thrings align to form a bound pair of thrings in a head-on arrangement, Fig.~\ref{fig:bound_pair}(d), that thwarts their individual locomotion.
See the supplemental material~\cite{SM} for a video of this simulation.
This dynamics is self-correcting as the outcome is generic and does not depend on the details of the initial condition, as long as there is a collision. However, in the absence of a collision,  each of the thrings continues its motion until it reaches the no-flux boundary of the medium, where it pairs with its mirror image. The formation of a bound state is an extremely robust process and represents a way to pin the centre of mass of the scroll ring, as an alternative to using solid obstacles~\cite{Steinbock:PRL:2009}.

\begin{figure}
\includegraphics[width=\columnwidth]{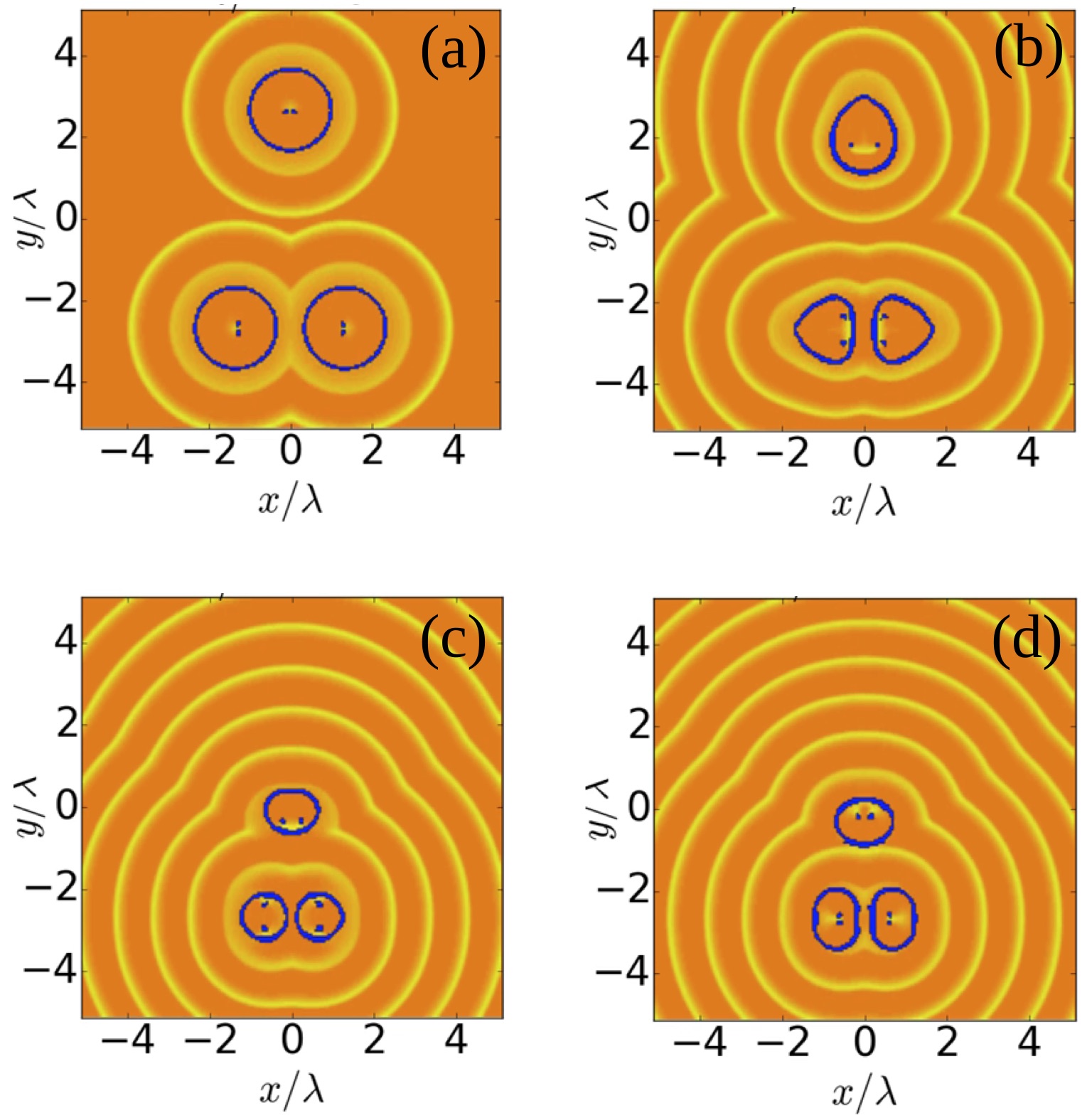}
\caption{The interaction of three thrings. (a) The situation shortly after simulated optical initialization ($t/T=2.8$), (b-c) the formation of a bound pair that is protected from invasion by an isolated thring ($t/T=15.8, \ 61.5$), (d) the long-term dynamics is an isolated thring trapped by the waves emanating from the bound pair ($t/T=92.7$).
\label{fig:third_wheel}
}
\end{figure}

Despite the attraction of a pair of thrings, the addition of a third thring does not lead to a triplet bound state, thus representing an example of the principle that more is different~\cite{Anderson:Science:1972}. This is demonstrated in
Fig.~\ref{fig:third_wheel}, where an isolated thring is initialized to swim towards a pair that will form a bound state. After the pair of thrings form a bound state, the frequency of the waves emitted by the pair is slightly higher than the waves produced by an isolated thring. This means that the wavefront collision interface slowly moves towards the isolated thring, until it is within a wavelength of the single thring. At this point, the isolated thring is no longer protected by its own shield of waves and is repeatedly slapped by the waves from the pair. The single thring is therefore swimming against the tide and this prevents it from joining the pair, consigning its fate to that of a permanent third-wheel, with the direction of its thwarted motion being normal to the wavefront emanating from the pair.
See the supplemental material~\cite{SM} for a video of this simulation.
As we demonstrate below, the increase in frequency of a pair makes them extremely robust against their environment and the advances of other thrings.

\begin{figure}
\includegraphics[width=\columnwidth]{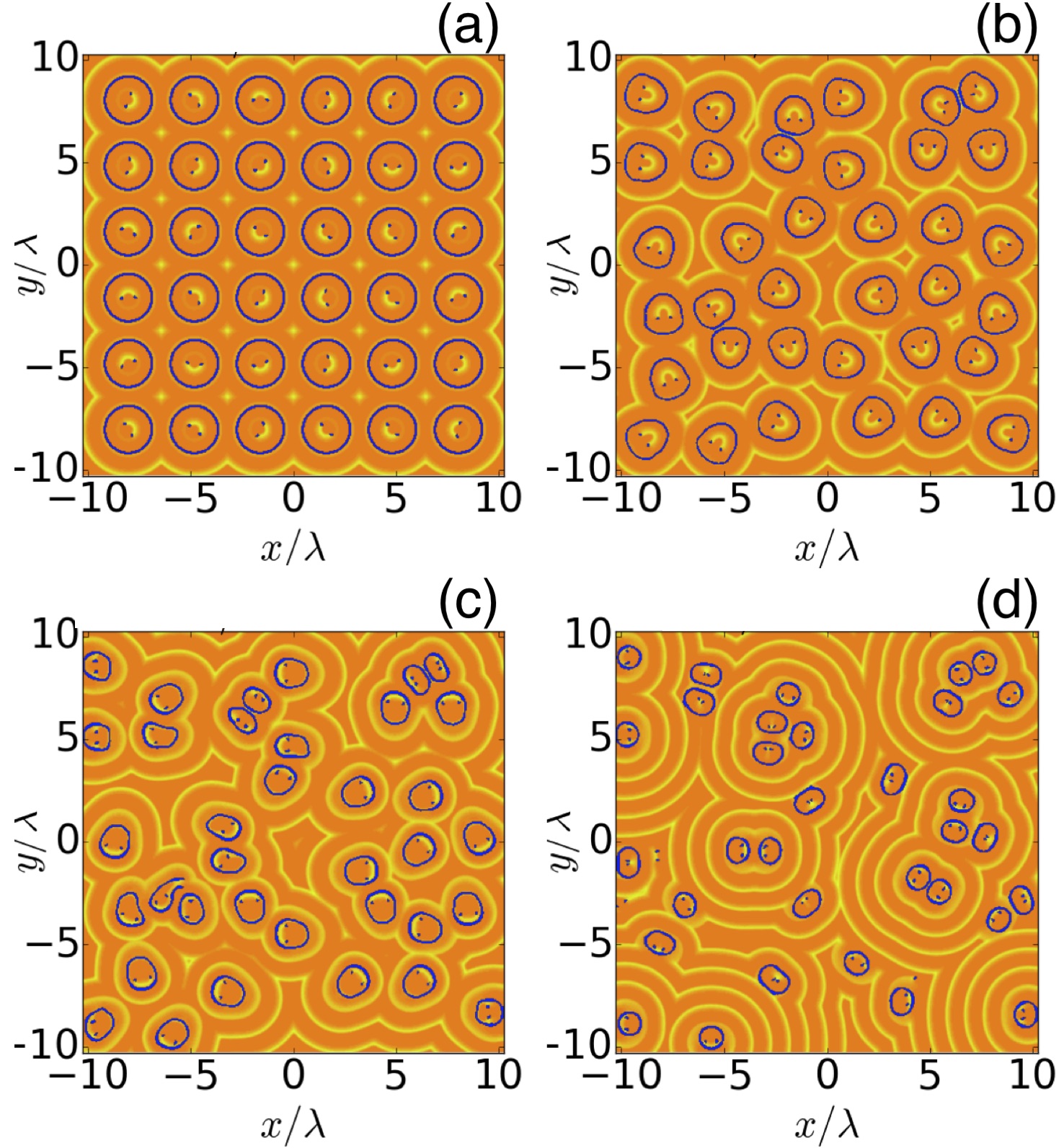}
\caption{The evolution of a colony containing 36 thrings. (a) The initial array of $6\times 6$ thrings initialized with a regular spacing but random swimming directions ($t/T=2.5$). (b) The early stages of the motion before the pairs are formed ($t/T=20.4$). (c) Collisions take place and pairs begin to form (including with mirror images at the no-flux boundary). A thring is destroyed  by a reconnection event ($t/T=34$). (d) Ultimately, a frozen state is reached, consisting of stable bound pairs together with single thrings trapped by the waves emitted from the pairs ($t/T=213$).
\label{fig:collective_behaviour}
}
\end{figure}

We have performed many simulations to test the hypothesis that thrings generically form robust compound pairs but not bound states of more than two thrings. In particular, we have tested this against the complicated spatio-temporal perturbations that are generated by many neighbouring thrings in a random colony.
A typical example is presented in Fig.~\ref{fig:collective_behaviour}, which illustrates the dynamics for a colony of 36 thrings. The initial positions of the thrings are regularly spaced in a $6\times 6$ grid, Fig.~\ref{fig:collective_behaviour}(a), however their initial swimming directions are randomized. In the early stages of the motion, Fig.~\ref{fig:collective_behaviour}(b), the thrings simply continue on the paths of the initial directions. As collisions take place, Fig.~\ref{fig:collective_behaviour}(c), several pairs are formed and locked in position. Note that in the bottom left corner of Fig.~\ref{fig:collective_behaviour}(c), a thring is being destroyed as it is squeezed out of shape by its neighbours. This results in a reconnection event, where the filament of the scroll ring contacts the surface of the thin excitable medium, breaking the closed loop and producing a contraction of the filament length until it disappears.  Ultimately, the system yields a surprising frozen state, Fig.~\ref{fig:collective_behaviour}(d), where all the surviving thrings, in this case 33 of the initial 36, have either formed robust bound state pairs (including pairing with a mirror image at the boundary of the medium), or are single thrings that are stuck swimming against the tide of higher frequency waves produced by one of the stationary pairs of thrings. 
See the supplemental material~\cite{SM} for a video of this simulation.

In summary, we have performed the first investigations of the dynamics, interaction and collective behaviour of a colony of thrings.
Rather than clustering, we find a very different type of self-organization for a colony of thrings, that results in a frozen state containing only bound pairs and isolated thrings, where all motion has stalled. As we have shown, this unusual state can be understood as the generic behaviour obtained from the application of the following simple rules of the game for thring interactions. Two adjacent thrings attract to form a fixed bound state, which then generates waves of a higher frequency than the waves produced by an isolated thring. The motion of a single thring stalls under the continued collision with higher frequency waves emitted from a pair that continually slap the filaments of the thring.
Single thrings remain unpaired, despite the existence of single neighbours, because they are trapped by an alignment to swim against the tide of waves from a bound state pair.

It is perhaps surprising that frozen states emerge as the generic outcome in this kind of nonequilibrium system, however, we have seen that the above simple rules provide an explanation of this phenomenon. Indeed it should be possible to predict the final frozen state given any initial system of thrings, using a simplified game of life type model, that includes a pairing rule for thrings and an alignment rule for isolated thrings given the nearest pair. More accurately, a more apt name would be a game of death model, as new thrings are never created but we have seen that existing thrings can occasionally be destroyed by collisions during the early stages of the evolution, before the pairs are established that ultimately freeze all motion.

A desired specific frozen state can be achieved by initially orientating thrings towards their intended partner. An example of this matchmaking process is provided in a video in the supplemental material~\cite{SM}, where 36 thrings have the same initial positions as in Fig.~\ref{fig:collective_behaviour}(a), but the orientations are chosen to create a symmetric frozen state, where all thrings survive to form a state containing only pairs, with no third-wheels. Explicitly, the orientations are chosen so that the thrings in rows one and five pair with the thrings directly below them in rows two and six, whereas the thrings in rows three and four pair with their neighbour to the left/right in the same row if they are in an even/odd column. Targeting symmetric pattern formation, as in this example, would be a convenient method to verify the theoretical results in future experiments.

The interaction of thrings represents an elegant and simplified platform for understanding the interactions between waves and filaments in excitable media more generally. It represents a quasi-two-dimensional arena in which to study filament motion induced by wave slapping due to differences in wave frequencies. This is a crucial mechanism that drives the more complicated motion of filament knots and links~\cite{Sutcliffe:PRE:2003,Sutcliffe:PRL:2016,Sutcliffe:PRE:2017,Sutcliffe:JPhysA:2018,Sutcliffe:Nonlinearity:2019,Gareth:PRE:2019} in a fully three-dimensional context. The work presented here provides a clear motivation to extend 
current experimental studies on thrings~\cite{Cincotti:arxiv:2019}, to advance capabilities that will provide experimental results for comparison to other systems displaying complex swimming dynamics and clustering, such as colloids, bacteria and nematic liquid shells~\cite{Maass:PRL:2019}, where chemical signalling and topology also play a promiment role.

 \section*{Acknowledgements}
This work is funded by the Leverhulme Trust Research Programme Grant No. RP2013-K-009, SPOCK: Scientific Properties of Complex Knots, F.M. acknowledges funding by the Danish National Research Foundation through a Niels Bohr Professorship to Thomas Pohl. Many thanks to Antonio Cincotti, Elizabeth Bromley, Andrew Lobb and Jonathan Steed for useful discussions. The computations were performed on Hamilton, the Durham University HPC cluster.

\end{document}